\documentstyle[12pt]{article}
\input epsf
\begin{document}
\begin{titlepage}
\title{ {\bf Exact Finite Size Results on}\\
 {\bf the Ising Model in 2D Curved
Space} }

\vspace{2cm}
\author{ {\bf J. Gonz\'alez\dag}\thanks{Electronic mail:
emgonzalez@iem.csic.es . Work partly supported by CICYT under contract
AEN90-0139.} \mbox{$\:$} and
{\bf M. A. Mart\'{\i}n-Delgado\ddag}\thanks{Electronic mail:
martind@puhep1.princeton.edu . Work partly supported by CICYT under
contract AEN90-0034.} \\
\mbox{}    \\
\dag{\em Instituto de Estructura de la Materia}\\
{\em Serrano 123, 28006-Madrid, Spain } \\
\ddag{\em Joseph Henry Laboratories}\\
{\em Princeton University}\\
{\em Princeton NJ 08540, USA} }
\vspace{5cm}
\date{}
\maketitle
\def\baselinestretch{1.3}
\begin{abstract}

We propose an approach to statistical systems on
lattices with sphere-like topology. Focusing on the Ising model,
we consider the thermodynamic limit along a sequence of lattices
which preserve the {\em fixed} large scale geometry. The
hypothesis of scaling appears to hold at criticality, pointing
at a sensible definition of the continuum limit of the model in
the curved space.

\end{abstract}

\vspace{2cm}

\vskip-17.0cm
\rightline{PUPT-1367}
\rightline{{\bf December 1992}}
\vskip3in
\end{titlepage}

\newpage
\def\baselinestretch{1.5}
\noindent
The elucidation of the effects of gravity on statistical systems
has already deserved a lot of attention. The discussion has been
mainly centered on the influence of quantizing the spatial
geometry of two-dimensional statistical models at criticality.
The determination of the gravitational dressing of conformal
weights can be considered one of the most important theoretical
breakthroughs of the past decade. Let us recall that it was
first achieved in the light-cone gauge formulation of quantum
gravity \cite{pol}, and later by use of conformal field theory
methods \cite{dk}. The
physical problem which is adressed by this study is that of
considering cooperative phenomena on a fluctuating surface, and
it is
in this kind of setting that the gravitational scaling dimensions
have been confirmed by the method of dynamical triangulation of
random surfaces \cite{rs,cont,kazakov}.
A different topic arises, however, by considering
the effect on the critical behaviour of a fixed spatial geometry.
In this  respect, it is important to stress that the effects of
curvature have not been assessed on statistical systems. Since
these are made of a discrete number of lattice points, the main
drawback in posing the problem concerns the way in which the
thermodynamic limit can  be approached keeping fixed the spatial
geometry. Focusing in two-dimensions, it is not obvious how to
enforce this constraint, specially when the mean curvature over
the surface is positive. This corresponds to having a positive
Euler number, which forces the  topology to be  that of the sphere.
On the other hand, this is the first case of interest since
quite significant physical  systems are already known which
realize the above  condition on the curvature, from a simplicial
point of view. The most notorious instance is the
buckminsterfullerene, the molecule with sixty carbon atoms
placed at the  vertices of a truncated icosahedron \cite{sci}. Recently,
carbon aggregates have been discovered in which  each shell has
an increasing number of atoms disposed on a closed lattice of
sphere-like  topology \cite{ug}. These kind of systems invite  to open the
study of statistical properties in the presence of nonvanishing
curvature.

{}From the theoretical point of view, the basic question is to
give a  growing pattern by which a lattice can be built
systematically from a previous one, keeping the same
distribution of the curvature on the new scale. In this letter
we  propose a definition of the thermodynamic limit along some
simple classes of two-dimenional lattices with the topology of
the sphere. According to simplicial geometry, one assigns to
every lattice face with  number of vertices $n_{i}$ the surface
element
\begin{equation}
\Delta S_{i} = \frac{1}{3} n_{i}
\end{equation}
and a value of the curvature given by
\begin{equation}
R_{i} = \pi \frac{6 - n_{i}}{n_{i}}
\end{equation}
The search for the mentioned growing pattern leads to consider
hexagonal lattices with a finite number of defects. Let us
constrain the problem by allowing only convex ($R_{i} > 0$) dislocations.
It turns out
that these have to be placed according to the spatial
orientation of the vertices  of the tetrahedron, the octahedron
or the icosahedron. To be more  precise, the lattices we
consider have as building blocks triangular pieces of honeycomb
lattice  of the type shown in Figure 1, which are to be
assembled as the  faces of the given regular polyhedron \cite{nos}. When
this is done, we end up with a closed lattice with constant
coordination number in which all the curvature is concentrated
at a finite number of locations. These correspond to four
three-fold rings in the case of the tetrahedron, six four-fold
rings in the octahedron and twelve five-fold rings in the icosahedron.
Given the polyhedron, though, the family of lattices of this
type embeded on its surface has an infinite number of members,
with their linear dimension increasing as an integer multiple of
the smallest one. In the case of the tetrahedron, for instance,
the first lattice has 12 points ---those of a truncated
tetrahedron---, the next has 48 points, and so on following the
general rule that the $N^{th}$ generation of the family has $12
N^{2}$ lattice points. The important point in this construction
is that for all the members of a given family the curvature is
distributed in the same way on the lattice. Actually, in the
thermodynamic limit we are dealing with a kind of discrete
version of an orbifold, that is, a degenerate manifold in which
all the curvature is concentrated at a finite number of
points ---the vertices of the  regular polyhedron.
When dealing with a given statistical model on these lattices,
it becomes  plausible the physical picture that changing from a
generation to another inside the same family amounts to consider
the same system, just with a different length scale. The
hypothesis of scaling is susceptible of being checked on
suitable observables, and  for this purpose precise measurements
are presented below in the  case of a simple system.

We focus in what follows on the Ising model formulated over the
family of lattices embeded on the tetrahedron. In  the case of
the Ising model we may take advantage of powerful techniques
developed for the computation of partition functions. We refer
to the dimer formulation of the problem, which we now briefly
review \cite{mont,mw}.
Given a collection of spins $\{ \sigma_{i} \}$, with $i$ running
over all the lattice points, the knowledge of the partition
function ${\cal Z}$ requires performing the sum over all
posible configurations
\begin{equation}
{\cal Z} = \sum_{\sigma_{i} = \pm 1} \mbox{\Large $e^{\beta
    \sum_{<i,j>} \sigma_{i} \sigma_{j}}$} \label{11}
\end{equation}
where the sum in the hamiltonian runs over all the lattice links
$\; <i,j>$.
The consideration of the high-temperature expansion about $\beta
= 0$ leads to the alternative expression
\begin{equation}
{\cal Z} = (cosh \: \beta)^{l} \; \sum_{\{ l_{i} \}} (tanh \: \beta)^{n_{i}}
                              \label{12}
\end{equation}
where $l$ is the total number of links, $\{ l_{i}
\}$ stands for the collection of all the closed nonintersecting
loops on the lattice and $\{ n_{i} \}$ is the respective number of links
for each of them. The dimer approach manages to compute this
sum by translating it into the evaluation of Pfaffians of
suitable antisymmetric operators. This is done as follows. One
has first to consider the decorated lattice, which in our case
is formed by inserting a triangle in place of each of the points
of the original lattice. A typical decorated lattice looks as
that depicted  in Figure 2, which corresponds to the lattice of
the second generation. Let us recall the definition of
elementary polygon and the notion of clockwise odd polygon in
the decorated lattice. Elementary polygons are those which do not
contain lattice points in their interior. An orientation may be
assigned to each of the links of the decorated lattice, which is
represented for practical purposes by drawing corresponding
arrows on them. A polygon is said to be clockwise odd when the
number of arrows pointing in the clockwise direction on the
polygon is odd. Obviously, there is always a manner of disposing
the arrows on the lattice as to render all the elementary
polygons clockwise odd. On a planar lattice all the work is done
by finding one of these systems of arrows. Kasteleyn theorem \cite{kast}
assures then that the combinatorics of the sum in (\ref{12}) can
be reproduced by computing the Pfaffian of a single
antisymmetric operator $A$. This is given by a coordination
matrix on the decorated lattice, such that $A_{ij}$ is positive
if an arrow goes from $i$ to $j$ and negative in the opposite
case. The $\beta$-dependence of the sum is fixed by taking the
absolute value of matrix elements for triangle links equal to $z
= tanh \: \beta$ (for $\beta > 0$), and the absolute value of the
rest equal to $1$.

Our lattices embedded on the tetrahedron fall into the category
of planar lattices, so that Kasteleyn theorem is at work. There
are some minor details concerning the possibility of defining an
order relation and the so-called transition cycles on our
lattices, which can be elaborated without much complication. The
sum of interest in (\ref{12}) can be represented by
the square root of the determinant of some antisymmetric
operator $A$ following the steps given above. There is, in fact,
a systematic way of forming a system of arrows making all
elementary polygons clockwise odd, generation after generation.
Figure 2 illustrates a particularly convenient way of
superposing the decorated lattices on the plane. The arrows on
the triangle links may all be chosen clockwise, while those
connecting triangles follow the regular pattern shown in the
bulk. There are only a reduced number of arrows along the
boundary which cannot conform to this rule and have to be fixed
to make all elementary polygons clockwise odd. In general,
progressing in the family of lattices from one generation to the
next amounts to add a row of hexagons at the top and another at
the bottom in the representation of Figure 2, expanding
accordingly the horizontal dimension. The proposed system of
arrows along the boundary follows also a regular pattern
beginning at each three-fold ring, which makes the construction
very simple for arbitrarily large lattices.

Once we have determined the system of arrows for a given
lattice, it only remains the technical problem of computing the
determinant of the corresponding antisymmetric operator $A$. The
partition function is given now by
\begin{equation}
{\cal Z} = (cosh \: \beta)^{l} \; (det \: A)^{1/2} \label{13}
\end{equation}
A direct symbolic computation of the determinant of $A$ cannot
be achieved in a reasonable amount of time beyond the first
generation, which already requires to handle a determinant of
order 36. We have deviced a trick, however, which makes possible
the exact computation of the partition function at least up to the  third
generation. The point is that for the kind of operators we are
dealing with $(det \: A)^{1/2}$ is a polynomial in  the variable
$z = tanh \: \beta$, whose determination demands only the knowledge
of a finite number of coefficients. For any of our lattices the
degree of the polynomial equals the total  number of lattice
points $v$. This can be checked by noticing that it always
exists a closed nonintersecting path which goes through all
the points in the lattice. Once it is known that the polynomial
$(det \: A)^{1/2}$ has degree $v$ for each  lattice, we only need to
perform a number of evaluations $v$ of this quantity at
different values of $z$ to identify all the coefficients
(the zeroth order term of the polynomial is obviously equal to
$1$). From a technical point of view, the coefficients are
integer numbers and can be computed without error by
sufficiently precise  evaluations of $(det \: A)^{1/2}$. We have
carried out this program up to the third generation of lattices
by using $Mathematica^{TM}$. In practice, what we have done at
each generation is to compute with full precision the square
root of the determinant at values $z = 1, 2, \ldots v$, for
which the result of the operation is itself an integer. Then by
solving a linear system the exact coefficients of the polynomial
are obtained. The partition function of the Ising model for the
first generation lattice of 12 points is, for instance,
\begin{equation}
{\cal Z} = \left( \frac{1}{1 - z^{2}} \right)^{9}
         (1 + 4z^{3} + 10z^{6} + 12z^{7} + 15z^{8} + 24z^{9} +
         30z^{10} + 24z^{11} + 8z^{12})   \label{14}
\end{equation}
For the lattice of the  second generation the partition function
turns out to be
\begin{eqnarray}
{\cal Z} & = & \left( \frac{1}{1 - z^{2}} \right)^{36}
                 (1   + 4  z^{3} + 28  z^{6} + 12  z^{7} +
               96  z^{9} + 96  z^{10} + 60  z^{11} + 356  z^{12}
                           \nonumber      \\
        &  &     +  576  z^{13} + 444  z^{14} + 1396  z^{15} +
               2814  z^{16} + 3456  z^{17} + 6488  z^{18}
                            \nonumber    \\
        &  &       +  12372  z^{19} + 18492  z^{20} + 29216  z^{21} +
               50028  z^{22} + 77292  z^{23}
                            \nonumber     \\
        &  &    +   115572  z^{24} + 177888  z^{25} + 265680  z^{26} +
               383812  z^{27} + 545436  z^{28}
                             \nonumber    \\
        &  &    +   760704  z^{29} + 1032180  z^{30} + 1356444  z^{31} +
               1728729  z^{32} + 2133696  z^{33}
                             \nonumber    \\
        &  &    +   2532312  z^{34} + 2870712  z^{35} + 3112708  z^{36} +
               3218976  z^{37} + 3142680  z^{38}
                             \nonumber    \\
        &  &    +   2875064  z^{39} + 2441460  z^{40} + 1898112  z^{41} +
               1322784  z^{42} + 799344  z^{43}
                              \nonumber   \\
        &  &    +   404928  z^{44} + 165888  z^{45} + 52272  z^{46} +
               12096  z^{47} + 1728  z^{48})      \label{15}
\end{eqnarray}
For these first two generations the alternative method of
counting polygonal paths on the lattice can be worked out,
producing the same respective high-temperature series as shown
above. Our computational approach  appears to be much more
efficient, however, starting with the lattice of the third
generation. After about 98 hours of CPU time on a Silicon
Graphics 4D/480S, we have
obtained the coefficients of the high-temperature series for the
lattice of 108 points. The expression for the partition function
is given in the Appendix. The
significance of having at our disposal the analytic expressions
of the partition functions is that they allow us to address some
questions with great precision. One of them refers to the location
of  the zeroes of the partition functions in the $z$ complex
plane. These are plotted, for the lattices of the second and
third generations, in Figures 3 and 4 respectively. The
frustration of the lattice clearly reflects in the lack of
symmetry of both partition functions under the exchange $\beta
\leftrightarrow -\beta$.
However, the
evolution of the zeroes in the complex plane points to the
formation of two accumulation points on the real axis in the
limit of large lattices, which signal the development of two
critical points respectively in the ferromagnetic and the
antiferromagnetic domain.

The inspection of the first derivatives of the free energy furnishes
also clear evidence of critical  behaviour in the
thermodynamic limit. In Figure 5 we show a plot of the second
derivative of the free energy with respect to $\beta$, for the
first three generations. At each generation we are able to
define a certain ``critical'' coupling constant, within each
domain, as that at which the specific heat reaches its maximum.
Let us focus, for instance, in the ferromagnetic  domain ($\beta
> 0$), where the critical behaviour has an  early development.
The critical  coupling constants for lattices up to the sixth
generation (432 points) are given in Table 1, along with the
corresponding values of the specific  heat at the  maximum.
The values  for the fourth, fifth and sixth generations result from
extensive numerical computations carried on the  representation
(\ref{13}) of the partition function. The analysis of finite
size effects allows to check if the scaling hypothesis applies
for the lattices in the curved space, being therefore helpful in
the computation of critical  exponents. Regarding the behaviour
of the finite size critical coupling  constant $\beta_{L}$, the
standard argument says that its difference with respect to the
value in the thermodynamic limit $\beta_{\infty}$ is fixed by
the point at which the correlation length $\xi$ reaches the
characteristic length dimension $L$ of the lattice. In terms of
the $\nu$ critical exponent we should have
\begin{equation}
\left| \beta_{L} - \beta_{\infty} \right| \sim L^{-1/\nu} \label{16}
\end{equation}
We give in Figure 6 the logarithmic plot of
$\: \beta_{L} - \beta_{\infty} \:$ versus the generation number
$N$, that is  also the appropriate length scale of the lattice.
We have assumed that $\beta_{\infty}$ coincides with the critical
coupling constant of the Ising model on a planar honeycomb
lattice, whose value is $\; \log(2 + \sqrt{3})/2 \approx 0.6585 $.
Quite remarkably, the correlation of the points in the plot is manifest,
including those corresponding to the smaller lattices. This
provides full support to the scaling law (\ref{16}). The linear
fit for the four bigger lattices in the plot gives a value
$1/\nu \approx  1.8$, with an uncertainty of about 10\%.
A significant drift from  this value should not be discarded for
bigger lattices, though the investigation of this point
requires finer computational techniques. A similar analysis can
be applied to the divergence of the  specific heat. Under the
assumption of scaling, the maximum value of the finite size
specific heat $C_{L}$ has to follow the law \cite{itzyk}
\begin{equation}
C_{L} \sim L^{-\alpha/\nu}  \label{17}
\end{equation}
A logarithmic plot of the  values for the specific heat in Table
 1  shows  again a proper linear correlation including the first
six generations. The exponent which arises from the linear fit
is, however,  rather small ($\sim 0.3 $ taking into account the
last four generations). The small number of lattices under
consideration makes impossible to discern whether the correct
critical behaviour follows a power law of the type (\ref{17}) or
is, in  fact, logarithmic. The progression to fairly  big
lattices may  be  the only way to shed  light on this question.

To summarize, we have proposed an approach to the introduction
of curvature on two-dimensional statistical systems in which the
size of the spatial lattice may be increased
keeping fixed the large scale geometry. We have developed it
studying the Ising model in particular, showing that the
hypothesis of scaling holds near the critical point. This
supports the point of view that the  thermodynamic limit taken
along our sequence of lattices provides a sensible definition of
the {\em continuum limit} of the Ising model on a
two-dimensional curved space. It would be nice to have an
alternative field theory  description of this limit, specially
if one were to confirm  unusual critical exponents for the
specific heat, the spontaneous magnetization or the
susceptibility of the model. The  precise determination of these
requires further work, as well as a different computational
approach. The study of finite size effects through the family of
lattices proposed appears to be worth not only at criticality,
but also to investigate possible integrability properties
off-criticality.

\vspace{2cm}

We are indebted to M. A. H. Vozmediano for her appreciated
contribution at the early stages of this work.
We want to thank also E. Marino, B. Nachtergaele and J. Salas
for useful comments. M. A. M-D. acknowledges financial support from a
postdoctoral fellowship of the Ministerio de
Educaci\'on y Ciencia (Spain).

\newpage
\section*{Appendix}
After the computation described in the text, the partition
function for the Ising model in the lattice of 108 points turns
out to be
\begin{eqnarray}
{\cal Z} &=&\left(\frac{1}{1 - z^{2}}\right)^{81}\:
(1+4z^{3}+58z^{6}+12z^{7}+216z^{9}+186z^{10}
                                               \nonumber \\
& &+60z^{11}+1601z^{12}+ 1296z^{13}+804z^{14}+5896z^{15}+10140z^{16}+7776z^{17}
                                                \nonumber \\
& &+32172z^{18}+51720z^{19}+59475z^{20}+138960z^{21}+287046z^{22}+375708z^{23}
                                                \nonumber     \\
& &+718858z^{24}+1381752z^{25}+2154546z^{26}+3696388z^{27}+6910026z^{28}
                                                \nonumber     \\
& &+11503728z^{29}+19206410z^{30}+34305804z^{31}+58642887z^{32}+98637960z^{33}
                                                \nonumber     \\
&
&+169526346z^{34}+290348280z^{35}+486916719z^{36}+822514464z^{37}+1387286346z^{38}
                                                \nonumber     \\
& &+2314063700z^{39}+3843366849z^{40}+6373069560z^{41}+10492090874z^{42}
                                                \nonumber     \\
& &+17155090536z^{43}+27935962830z^{44}+45201023376z^{45}+72645985740z^{46}
                                                \nonumber     \\
& &+116018542776z^{47}+184128512464z^{48}+290116289952z^{49}+453866505468z^{50}
                                                \nonumber     \\
& &+704934164280z^{51}+1086527017686z^{52}+1661488438896z^{53}+
2520032972416z^{54}
                                                \nonumber     \\
&
&+3790362808512z^{55}+5651112933240z^{56}+8349602813680z^{57}+12221741958408z^{58}
                                                \nonumber     \\
& &+17717112551232z^{59}+25427112163896z^{60}+36115134719952z^{61}
                                                \nonumber     \\
& &+50748717022488z^{62}+70522684237440z^{63}+96880972147407z^{64}
                                                \nonumber     \\
& &+131514489830256z^{65}+176340456853560z^{66}+233441767931004z^{67}
                                                \nonumber     \\
& &+304963031957238z^{68}+392959853131920z^{69}+499172664304386z^{70}
                                                \nonumber     \\
& &+624768901981884z^{71}+770015161230510z^{72}+933949132164936z^{73}
                                                \nonumber     \\
& &+1114059041938386z^{74}+1306018538801380z^{75}+1503582518119761z^{76}
                                                \nonumber     \\
& &+1698618381066144z^{77}+1881426714304464z^{78}+2041294948806144z^{79}
                                                \nonumber     \\
& &+2167319364960132z^{80}+2249464698097728z^{81}+2279669302012248z^{82}
                                                \nonumber     \\
& &+2252977033783488z^{83}+2168397154335735z^{84}+2029388221287936z^{85}
                                                \nonumber     \\
& &+1843829681848794z^{86}+1623357219696420z^{87}+1382202439093638z^{88}
                                                \nonumber     \\
& &+1135591499049048z^{89}+898008996015174z^{90}+681607478672604z^{91}
                                                \nonumber     \\
& &+495011279161350z^{92}+342753592928640z^{93}+225365558696478z^{94}
                                                \nonumber     \\
& &+140070377148084z^{95}+81866387456505z^{96}+44729954075592z^{97}
                                                \nonumber     \\
& &+22694312231550z^{98}+10610393197704z^{99}+4531053196539z^{100}
                                                \nonumber     \\
& &+1749121091856z^{101}+ 602700352218z^{102}+182520341604z^{103}
                                                \nonumber     \\
& &+47584638729z^{104}+10387733544z^{105}+1824896250z^{106}
                                                \nonumber     \\
& &+237880152z^{107}+21484952z^{108})
\end{eqnarray}

\newpage
\begin{table}[p]
\centering
\begin{tabular}{|l|c|c|}  \hline \hline
  $N$    & $\beta_{L}$   &   $C_{L}$  \\  \hline \hline
 $ 1$    &  $0.4673(1) $    &  $1.4392(1)   $  \\ \hline
 $ 2$    &     $0.6082(1) $ &  $ 1.6520(1)  $  \\  \hline
  $3 $   &   $ 0.6342(1)$  &  $  1.8924(1)$  \\  \hline
 $4 $    &  $0.6439(1)$     &  $2.0798(1)$     \\ \hline
$5$      &  $0.6486(1)$     &   $2.2308(1)$   \\ \hline
$6 $     & $0.6513(1)$    &  $2.3568(1)$    \\ \hline \hline
\end{tabular}
\caption{Respective  values of the coupling constant $\beta_{L}$
at which the specific heat reaches the maximum values $C_{L}$ for
the first six generations.}
\end{table}

\newpage
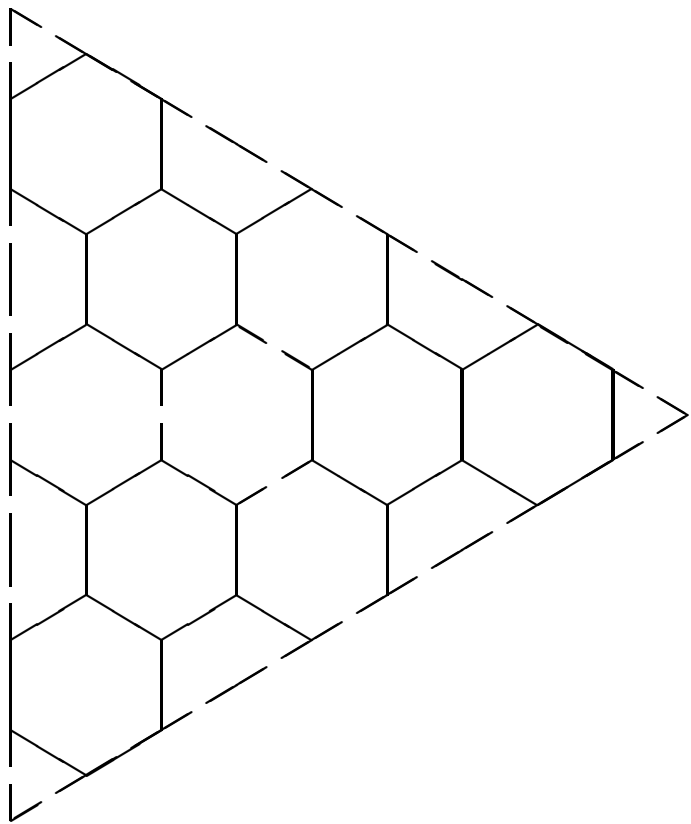
\begin{figure}[p]
\centering
\setlength{\unitlength}{1.2cm}
\begin{picture}(15,13)(2,0)

\thicklines

\multiput(5.66,2)(0,3){3}{\line(5,-3){0.83}}
\multiput(6.5,1.5)(0,3){3}{\line(5,3){0.83}}
\multiput(7.33,5)(0,3){2}{\line(5,-3){0.83}}
\put(8.16,7.5){\line(5,3){0.83}}
\put(9,5){\line(5,-3){0.83}}
\put(9.83,4.5){\line(5,3){0.83}}
\put(10.66,5){\line(5,-3){0.83}}
\put(11.5,4.5){\line(5,3){0.83}}

\multiput(5.66,3)(0,3){3}{\line(5,3){0.83}}
\multiput(6.5,3.5)(0,3){3}{\line(5,-3){0.83}}
\multiput(7.33,3)(0,3){2}{\line(5,3){0.83}}
\put(8.16,3.5){\line(5,-3){0.83}}

\put(9,6){\line(5,3){0.83}}
\put(9.83,6.5){\line(5,-3){0.83}}
\put(10.66,6){\line(5,3){0.83}}
\put(11.5,6.5){\line(5,-3){0.83}}

\multiput(6.5,3.5)(0,3){2}{\line(0,1){1}}
\multiput(8.16,3.5)(0,3){2}{\line(0,1){1}}
\multiput(9.83,3.5)(0,3){2}{\line(0,1){1}}

\multiput(5.66,2)(0,6){2}{\line(0,1){1}}
\multiput(7.33,2)(0,6){2}{\line(0,1){1}}
\put(9,5){\line(0,1){1}}
\put(10.66,5){\line(0,1){1}}
\put(12.33,5){\line(0,1){1}}
\multiput(5.66,1)(0,1){9}{\line(0,1){0.4}}
\multiput(5.66,1.6)(0,1){9}{\line(0,1){0.4}}

\multiput(13.16,5.5)(-0.833,-0.5){9}{\line(-5,-3){0.333}}
\multiput(12.66,5.2)(-0.833,-0.5){9}{\line(-5,-3){0.333}}
\multiput(13.16,5.5)(-0.833,0.5){9}{\line(-5,3){0.333}}
\multiput(12.66,5.8)(-0.833,0.5){9}{\line(-5,3){0.333}}
\put(7.33,5){\line(0,1){0.4}}
\put(7.33,6){\line(0,-1){0.4}}
\put(8.16,4.5){\line(5,3){0.333}}
\put(9,5){\line(-5,-3){0.333}}
\put(8.16,6.5){\line(5,-3){0.333}}
\put(9,6){\line(-5,3){0.333}}
%

\end{picture}
\caption{Generic triangular block for honeycomb lattices embeded on the
tetrahedron, the octaedron and the icosahedron.}
\end{figure}

\newpage
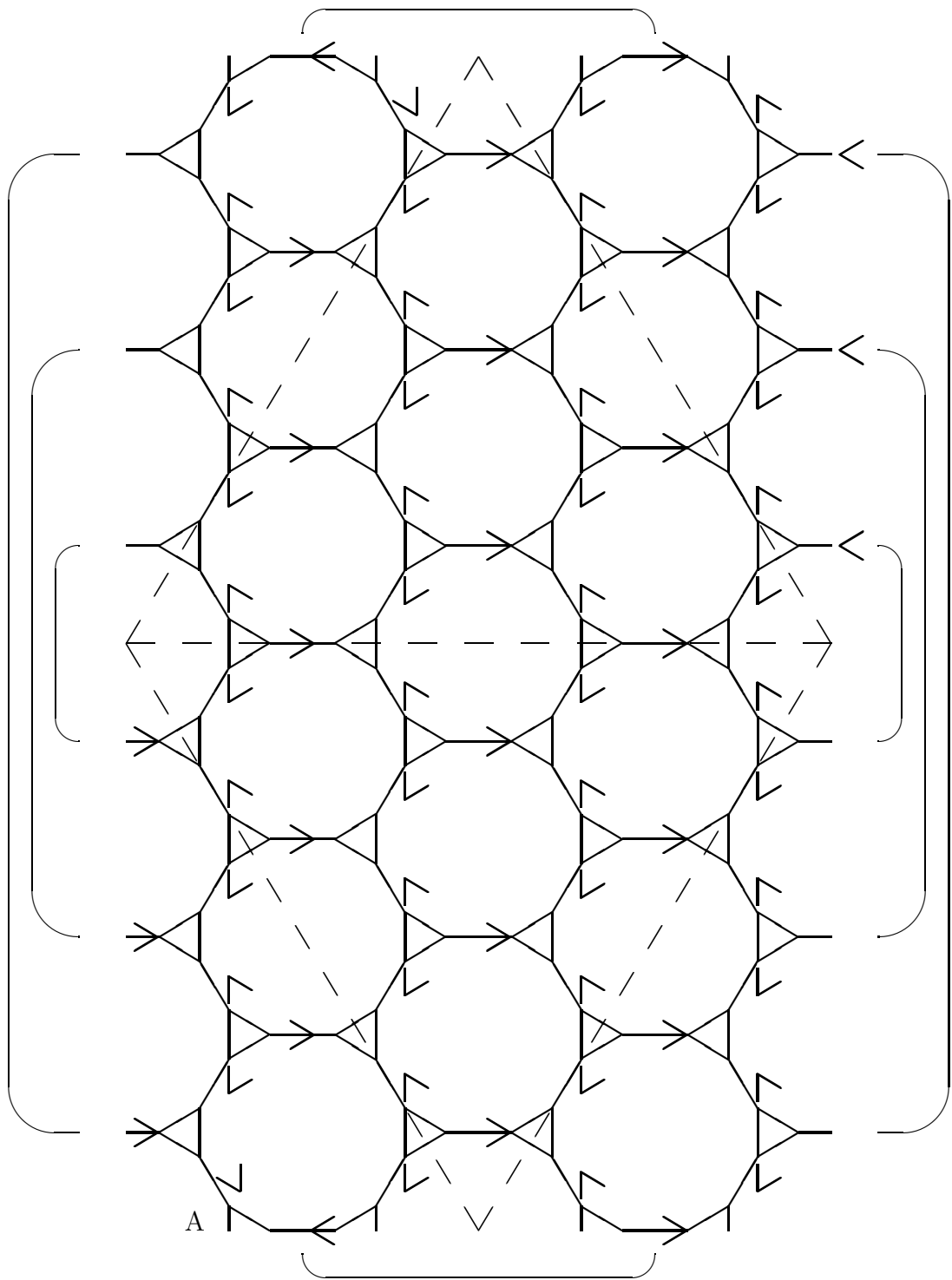
\begin{figure}[p]
\centering
\setlength{\unitlength}{1.8cm}
\begin{picture}(5,9.5)(-0.3,-1)

\thinlines
\put(5.9,5){\oval(0.4,1.6667)[r]}
\put(5.9,5){\oval(0.8,5)[r]}
\put(5.9,5){\oval(1.2,8.333)[r]}
\put(-0.9,5){\oval(0.4,1.6667)[l]}
\put(-0.9,5){\oval(0.8,5)[l]}
\put(-0.9,5){\oval(1.2,8.333)[l]}
\put(2.5,-0.2){\oval(3,0.4)[b]}
\put(2.5,10.2){\oval(3,0.4)[t]}
\multiput(-0.5,5)(0.48,0){13}{\line(1,0){0.24}}
\multiput(-0.5,5)(0.24,0.4){13}{\line(3,5){0.12}}
\multiput(-0.5,5)(0.24,-0.4){13}{\line(3,-5){0.12}}
\multiput(2.5,0)(0.24,0.4){13}{\line(3,5){0.12}}
\multiput(2.5,10)(0.24,-0.4){13}{\line(3,-5){0.12}}

\thicklines
\newsavebox{\arrr}
\savebox{\arrr}(1,1)[bl]{\begin{picture}(1,1)
\put(0,0){\line(-5,3){0.2}}
\put(0,0){\line(-5,-3){0.2}}
\end{picture}}
\newsavebox{\arrl}
\savebox{\arrl}(1,1)[bl]{\begin{picture}(1,1)
\put(0,0){\line(5,3){0.2}}
\put(0,0){\line(5,-3){0.2}}
\end{picture}}

\newsavebox{\arrul}
\savebox{\arrul}(1,1)[bl]{\begin{picture}(1,1)
\put(0,0){\line(5,-3){0.2}}
\put(0,0){\line(0,-1){0.2332}}
\end{picture}}
\newsavebox{\arrdl}
\savebox{\arrdl}(1,1)[bl]{\begin{picture}(1,1)
\put(0,0){\line(5,3){0.2}}
\put(0,0){\line(0,1){0.2332}}
\end{picture}}

\newsavebox{\arrdr}
\savebox{\arrdr}(1,1)[bl]{\begin{picture}(1,1)
\put(0,0){\line(-5,3){0.2}}
\put(0,0){\line(0,1){0.2332}}
\end{picture}}

\newsavebox{\toy}
\savebox{\toy}(3,2)[bl]{\begin{picture}(3,2)
\put(1.625,-0.2083){\line(0,1){0.4167}}
\put(1.278,0.0){\line(5,3){0.3472}}
\put(1.278,0.0){\line(5,-3){0.3472}}
\put(0.7222,0.0){\line(1,0){0.5556}}
\put(0.375,0.2083){\line(5,-3){0.3472}}
\put(0.375,-0.2083){\line(5,3){0.3472}}
\put(0.375,-0.2083){\line(0,1){0.4167}}
\put(0.375,0.2083){\line(-3,5){0.25}}
\put(1.625,0.2083){\line(3,5){0.25}}
\end{picture}}

\newsavebox{\cort}
\savebox{\cort}(3,2)[bl]{\begin{picture}(3,2)
\put(1.625,0.0){\line(0,1){0.20835}}
\put(1.278,0.0){\line(5,3){0.3472}}

\put(0.7222,0.0){\line(1,0){0.5556}}
\put(0.375,0.2083){\line(5,-3){0.3472}}

\put(0.375,0.0){\line(0,1){0.20835}}
\put(0.375,0.2083){\line(-3,5){0.25}}
\put(1.625,0.2083){\line(3,5){0.25}}
\end{picture}}

\newsavebox{\two}
\savebox{\two}(3,2)[bl]{\begin{picture}(3,2)
\put(1.625,-0.2083){\line(0,1){0.20835}}

\put(1.278,0.0){\line(5,-3){0.3472}}
\put(0.7222,0.0){\line(1,0){0.5556}}

\put(0.375,-0.2083){\line(5,3){0.3472}}
\put(0.375,-0.2083){\line(0,1){0.20835}}
\end{picture}}
\newsavebox{\arr}
\savebox{\arr}(3,2)[bl]{\begin{picture}(3,2)
\put(-0.125,0.2083){\line(5,-3){0.3472}}
\put(-0.125,-0.2083){\line(5,3){0.3472}}
\put(-0.125,-0.2083){\line(0,1){0.4167}}
\put(-0.125,0.2083){\line(-3,5){0.25}}
\put(0.2222,0.0){\line(1,0){0.2778}}
\end{picture}}
\newsavebox{\abj}
\savebox{\abj}(3,2)[bl]{\begin{picture}(3,2)
\put(0.125,-0.2083){\line(0,1){0.4167}}
\put(-0.222,0.0){\line(5,3){0.3472}}
\put(-0.222,0.0){\line(5,-3){0.3472}}
\put(0.125,0.2083){\line(3,5){0.25}}
\put(-0.5,0.0){\line(1,0){0.2778}}
\end{picture}}
\put(0,0){\usebox{\cort}}
\multiput(0,1.6667)(0,1.6667){5}{\usebox{\toy}}
\multiput(1.5,0.8333)(0,1.6667){6}{\usebox{\toy}}
\put(3.0,0.0){\usebox{\cort}}
\multiput(3.0,1.6667)(0,1.6667){5}{\usebox{\toy}}
\multiput(0.0,0.8333)(0,1.6667){6}{\usebox{\abj}}
\multiput(5.0,0.8333)(0,1.6667){6}{\usebox{\arr}}
\put(0.0,10.0){\usebox{\two}}
\put(3.0,10.0){\usebox{\two}}

\multiput(1.1,1.666667)(0,1.666667){5}{\usebox{\arrr}}
A
\multiput(2.6,0.83333)(0,1.666667){6}{\usebox{\arrr}}
\multiput(4.1,0.0)(0,1.666667){7}{\usebox{\arrr}}
\multiput(-0.4,0.83333)(0,1.666667){3}{\usebox{\arrr}}
\multiput(0.9,0.0)(0,10.0){2}{\usebox{\arrl}}
\multiput(5.4,5.83333)(0,1.666667){3}{\usebox{\arrl}}
\multiput(0.2,2.1666667)(0,1.666667){5}{\usebox{\arrul}}
\multiput(1.7,1.33333)(0,1.666667){5}{\usebox{\arrul}}
\multiput(3.2,0.5)(0,1.666667){6}{\usebox{\arrul}}
\multiput(4.7,1.33333)(0,1.666667){6}{\usebox{\arrul}}
\multiput(0.2,1.1666667)(0,1.666667){6}{\usebox{\arrdl}}
\multiput(1.7,0.33333)(0,1.666667){6}{\usebox{\arrdl}}
\multiput(3.2,1.1666667)(0,1.666667){6}{\usebox{\arrdl}}
\multiput(4.7,0.33333)(0,1.666667){6}{\usebox{\arrdl}}
\put(0.3,0.333333){\usebox{\arrdr}}
\put(1.8,9.5){\usebox{\arrdr}}

%

\end{picture}
\caption{Decorated lattice for the second generation. The outer
lines show the identifications of boundary links which embed the lattice
on the tetrahedron.}

\end{figure}

\newpage
\begin{figure}[p]
\centering
\setlength{\unitlength}{1cm}
\begin{picture}(14.2,17)
\put(0.8,1.5){\makebox(12.8,17.85)}
%

\centerline{\epsfysize 8.5 in \epsfbox{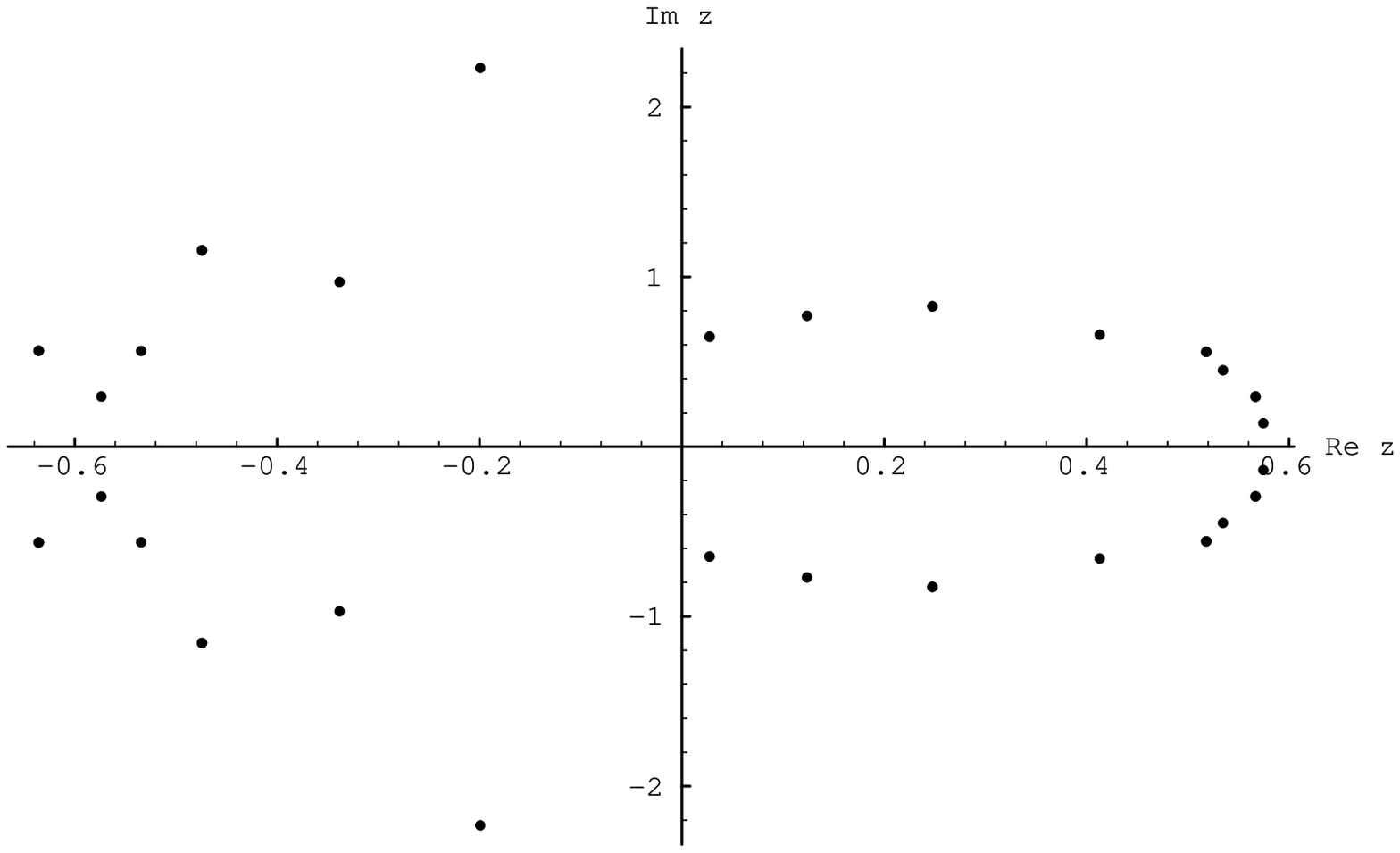}}

\end{picture}
\caption{Plot in the $z$-complex plane of the zeros of the
partition function for the second generation honeycomb lattice
on the tetrahedron.}
\end{figure}

\newpage
\begin{figure}[p]
\centering
\setlength{\unitlength}{1cm}
\begin{picture}(14.2,17)
\put(0.8,1.5){\makebox(12.8,17.85)}
%

\centerline{\epsfysize 8.5 in \epsfbox{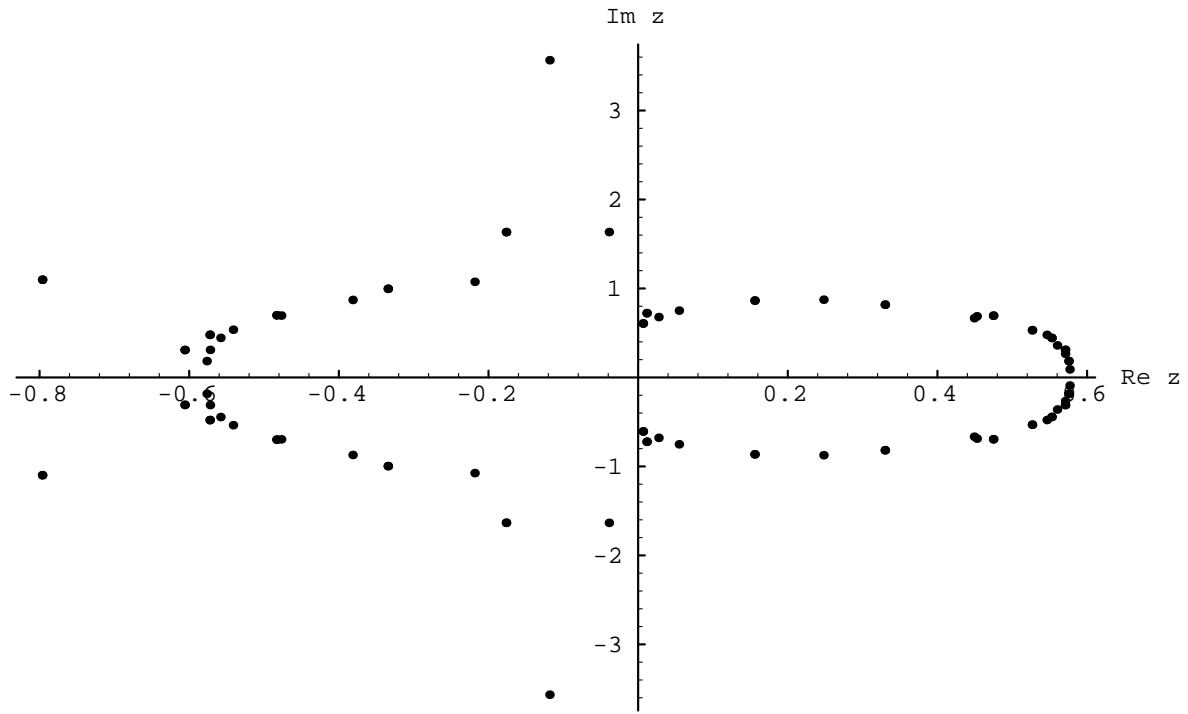}}

\end{picture}
\caption{Plot in the $z$-complex plane of the zeros of the
partition function for the third generation honeycomb lattice
on the tetrahedron.}
\end{figure}

\newpage
\begin{figure}[p]
\centering
\setlength{\unitlength}{1cm}
\begin{picture}(14.2,17)
\put(0.8,1.5){\makebox(12.8,17.85)}
%

\centerline{\epsfysize 8.5 in \epsfbox{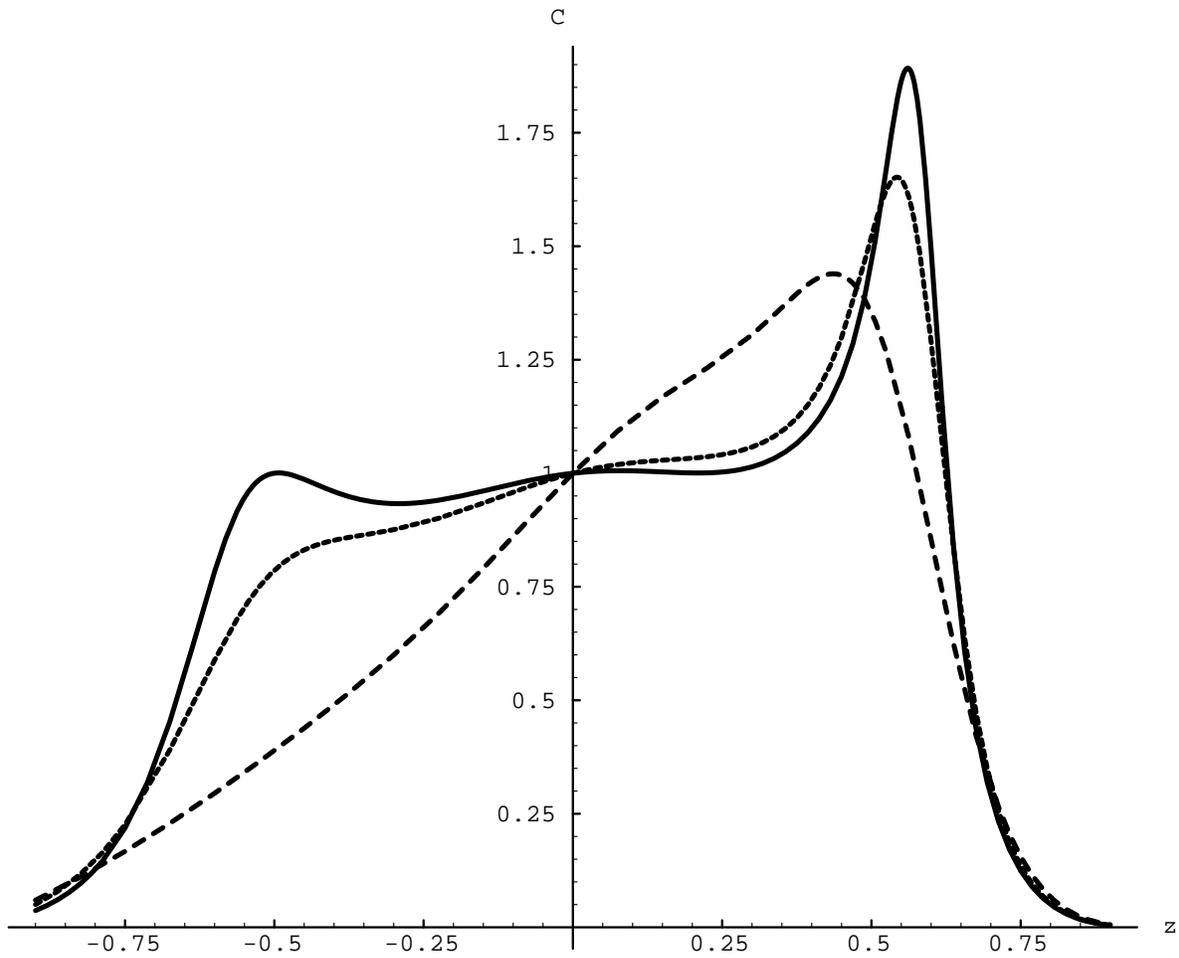}}

\end{picture}
\caption{Representation of the second derivative of the free
energy with respect to $\beta$ versus $z = tanh \:\beta$ for
the lattices of 12 points (long--dash line), 48 points (short--dash
line) and 108 points (full line).}
\end{figure}

\newpage
\begin{figure}[p]
\centering
\setlength{\unitlength}{1cm}
 \begin{picture}(14.2,17)
\put(0.8,3.5){\makebox(12.8,15)}
%

\centerline{\epsfysize 8.5 in \epsfbox{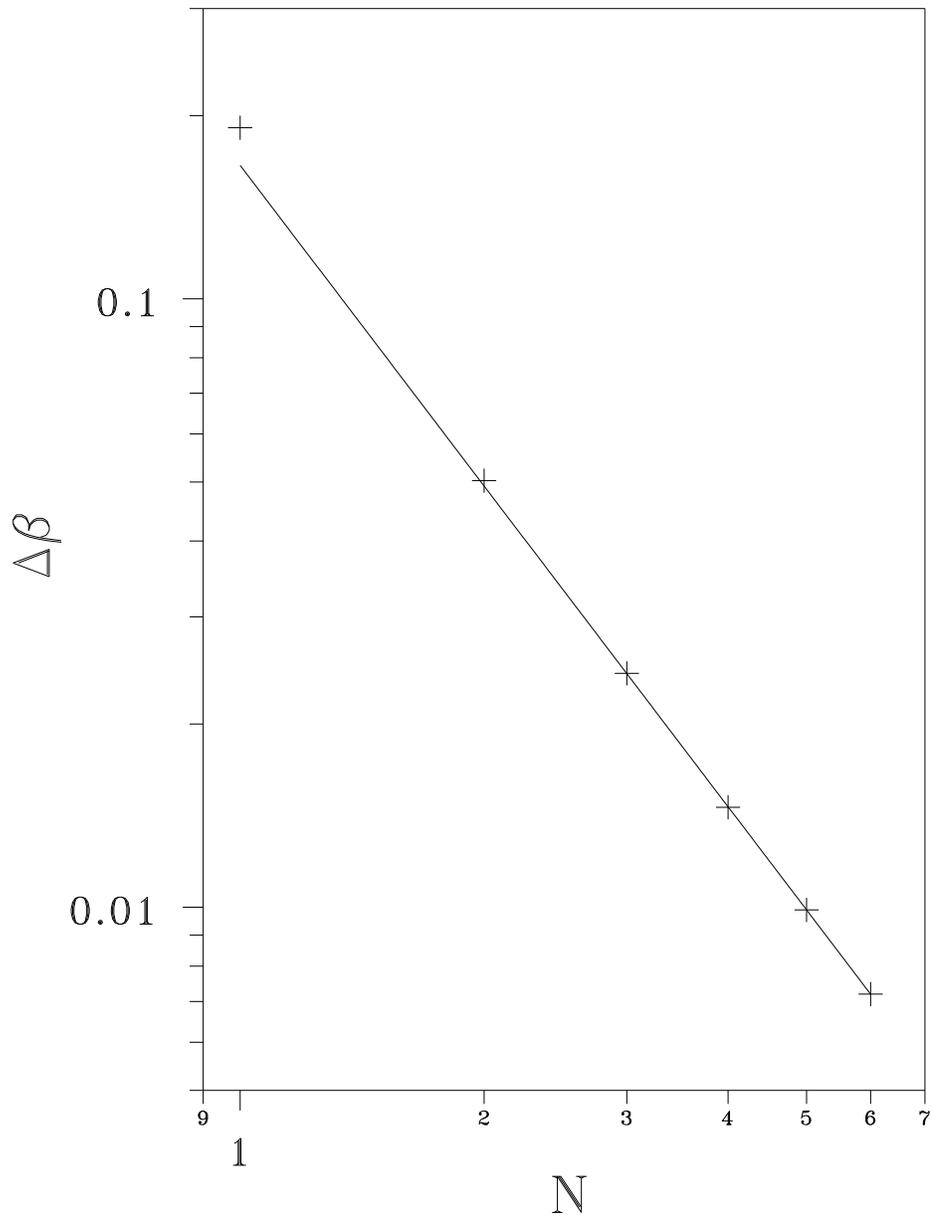}}

\end{picture}
\caption{Logarithmic plot of $\beta_{L} - \beta_{\infty}$ versus
the linear dimension $L$ for the lattices of the first six
generations. The straight line corresponds to the linear fit for
the last four points.}
\end{figure}

 \end{document}